\documentclass[12pt]{article}
\usepackage{amsmath,amssymb,rotating,subfigure,color,fullpage}

\newdimen \myunit
\newdimen \myhsize
\newdimen \myvsize
\newcommand{\wire}[1]{\raisebox{3\myunit}[0cm][0cm]{\small #1}}
\newcommand{\smwidth}{8}
\newcommand{\smhalf}{4}
\newcommand{\qthicklines}{\linethickness{1.5\myunit}}

\newcommand{\qheight}{10}
\newcommand{\qtopheight}{5}
\newcommand{\qtopheightminusthreehalves}{3.5}
\newcommand{\qtopheightplustwo}{7}
\newcommand{\qbottomheight}{5}
\newcommand{\qbottomheightminusthreehalves}{3.5}
\newcommand{\qbottomheightplustwo}{7}

\definecolor{darkgreen}{rgb}{0,0.5,0}

\def\cn#1{
\begin{picture}(4,\qheight)(0,0)
  \put(2,\qtopheight){\circle*{2}}
  \put(0,\qtopheight){\line(1,0){4}}
\ifcase #1
    \put(2,0){\line(0,1){\qheight}}
\or \put(2,\qtopheight){\line(0,1){\qbottomheight}}
\or \put(2,\qtopheight){\line(0,-1){\qtopheight}}
\fi
\end{picture}
}

\def\ccn#1#2{
\begin{picture}(4,\qheight)(0,0)
  \put(0,\qtopheight){\line(1,0){4}}
\textcolor{#1}{
  \put(2,\qtopheight){\circle*{2}}
\ifcase #2
    \put(2,0){\line(0,1){\qheight}}
\or \put(2,\qtopheight){\line(0,1){\qbottomheight}}
\or \put(2,\qtopheight){\line(0,-1){\qtopheight}}
\fi}
\end{picture}
}

\def\nn#1{
\begin{picture}(4,\qheight)(0,0)
  \put(2,\qtopheight){\circle{3}}
  \put(0,\qtopheight){\line(1,0){.5}}
  \put(4,\qtopheight){\line(-1,0){.5}}
\ifcase #1
    \put(2,0){\line(0,1){\qtopheightminusthreehalves}}
    \put(2,\qheight){\line(0,-1){\qbottomheightminusthreehalves}}
\or \put(2,\qheight){\line(0,-1){\qbottomheightminusthreehalves}}
\or \put(2,0){\line(0,1){\qtopheightminusthreehalves}}
\fi
\end{picture}
}

\def\cnn#1#2{
\begin{picture}(4,\qheight)(0,0)
  \textcolor{#1}{\put(2,\qtopheight){\circle{3}}}
  \put(0,\qtopheight){\line(1,0){.5}}
  \put(4,\qtopheight){\line(-1,0){.5}}
\textcolor{#1}{
\ifcase #2
    \put(2,0){\line(0,1){\qtopheightminusthreehalves}}
    \put(2,\qheight){\line(0,-1){\qbottomheightminusthreehalves}}
\or \put(2,\qheight){\line(0,-1){\qbottomheightminusthreehalves}}
\or \put(2,0){\line(0,1){\qtopheightminusthreehalves}}
\fi}
\end{picture}
}

\def\xn#1{
\begin{picture}(4,\qheight)(0,0)
  \put(2,\qtopheight){\circle{4}}
  \put(0,\qtopheight){\line(1,0){4}}
\ifcase #1
    \put(2,0){\line(0,1){\qheight}}
\or \put(2,\qheight){\line(0,-1){\qbottomheightplustwo}}
\or \put(2,0){\line(0,1){\qtopheightplustwo}}
\or \put(2,\qtopheightplustwo){\line(0,-1){4}}
\fi
\end{picture}
}

\def\cxn#1#2{
\begin{picture}(4,\qheight)(0,0)
  \textcolor{#1}{
  \put(2,\qtopheight){\circle{4}}
  \put(0,\qtopheight){\line(1,0){4}}
\ifcase #2
    \put(2,0){\line(0,1){\qheight}}
\or \put(2,\qheight){\line(0,-1){\qbottomheightplustwo}}
\or \put(2,0){\line(0,1){\qtopheightplustwo}}
\or \put(2,\qtopheightplustwo){\line(0,-1){4}}
\fi}
\end{picture}
}

\newcommand{\sn}[1]{
\begin{picture}(4,\qheight)(0,0)
\ifcase #1
    \put(0,\qtopheight){\line(1,0){4}}
\or \put(2,0){\line(0,1){\qheight}}
\or \put(0,\qtopheight){\line(1,0){4}}
    \put(2,0){\line(0,1){\qheight}}
\or \put(0,\qtopheight){\line(1,0){4}}
    \multiput(2,1)(0,\qtopheight){2}{\line(0,1){\qtopheightminusthreehalves}}
\fi
\end{picture}
}

\newcommand{\csn}[2]{
\begin{picture}(4,\qheight)(0,0)
\ifcase #2
    \put(0,\qtopheight){\line(1,0){4}}
\or \textcolor{#1}{\put(2,0){\line(0,1){\qheight}}}
\or \put(0,\qtopheight){\line(1,0){4}}
    \textcolor{#1}{\put(2,0){\line(0,1){\qheight}}}
\fi
\end{picture}
}

\def\sm#1{
\begin{picture}(\smwidth,\qheight)(0,0)
\ifcase #1
    \put(0,\qtopheight){\line(1,0){\smwidth}}
\or \put(\smhalf,0){\line(0,1){\qheight}}
\or \put(0,\qtopheight){\line(1,0){\smwidth}}
    \put(\smhalf,0){\line(0,1){\qheight}}
\or \put(0,\qtopheight){\line(1,0){\smwidth}}
    \multiput(\smhalf,1)(0,\qtopheight){2}{\line(0,1){\qtopheightminusthreehalves}}
\fi
\end{picture}
}

\def\sx#1{
\begin{picture}(12,\qheight)(0,0)
\ifcase #1
    \put(0,\qtopheight){\line(1,0){12}}
\or \put(6,0){\line(0,1){\qheight}}
\or \put(0,\qtopheight){\line(1,0){12}}
    \put(6,0){\line(0,1){\qheight}}
\or \multiput(0,\qtopheight)(2.6,0){\qtopheight}{\line(1,0){1.6}}
    \put(1,0){\line(0,1){\qheight}}
    \qthicklines
    \put(11,0){\line(0,1){\qheight}}
    \thinlines
\or \multiput(0,\qtopheight)(2.6,0){\qtopheight}{\line(1,0){1.6}}
    \qthicklines
    \put(1,0){\line(0,1){\qheight}}
    \thinlines
    \put(11,0){\line(0,1){\qheight}}
\fi
\end{picture}
}

\def\dx#1{
\begin{picture}(12,\qheight)(0,0)
  \put(6,\qtopheight){\circle*{2}}
  \put(0,\qtopheight){\line(1,0){12}}
\ifcase #1
    \put(6,0){\line(0,1){\qheight}}
\or \put(6,\qtopheight){\line(0,1){\qbottomheight}}
\or \put(6,\qtopheight){\line(0,-1){\qtopheight}}
\or {}
\or \put(1,0){\line(0,1){\qheight}}
    \qthicklines
    \put(11,0){\line(0,1){\qheight}}
    \thinlines
\or \qthicklines
    \put(1,0){\line(0,1){\qheight}}
    \thinlines
    \put(11,0){\line(0,1){\qheight}}
\fi
\end{picture}
}

\def\ex#1{
\begin{picture}(12,\qheight)(0,0)
  \put(6,\qtopheight){\circle{3}}
  \put(0,\qtopheight){\line(1,0){4.5}}
  \put(12,\qtopheight){\line(-1,0){4.5}}
\ifcase #1
    \put(6,0){\line(0,1){\qtopheightminusthreehalves}}
    \put(6,\qheight){\line(0,-1){\qbottomheightminusthreehalves}}
\or \put(6,\qheight){\line(0,-1){\qbottomheightminusthreehalves}}
\or \put(6,0){\line(0,1){\qtopheightminusthreehalves}}
\or {}
\or \put(1,0){\line(0,1){\qheight}}
    \qthicklines
    \put(11,0){\line(0,1){\qheight}}
    \thinlines
\or \qthicklines
    \put(1,0){\line(0,1){\qheight}}
    \thinlines
    \put(11,0){\line(0,1){\qheight}}
\fi
\end{picture}
}

\def\nt#1{
\begin{picture}(12,\qheight)(0,0)
  \put(6,\qtopheight){\circle{4}}
  \put(0,\qtopheight){\line(1,0){12}}
\ifcase #1
    \put(6,0){\line(0,1){\qheight}}
\or \put(6,\qheight){\line(0,-1){\qbottomheightplustwo}}
\or \put(6,0){\line(0,1){\qtopheightplustwo}}
\or \put(6,\qtopheightplustwo){\line(0,-1){4}}
\or \put(1,0){\line(0,1){\qheight}}
    \qthicklines
    \put(11,0){\line(0,1){\qheight}}
    \thinlines
    \put(6,\qtopheightplustwo){\line(0,-1){4}}
\or \qthicklines
    \put(1,0){\line(0,1){\qheight}}
    \thinlines
    \put(11,0){\line(0,1){\qheight}}
    \put(6,\qtopheightplustwo){\line(0,-1){4}}
\fi
\end{picture}
}

\def\ox#1{
\begin{picture}(12,\qheight)(0,0)
  \put(6,\qtopheight){\circle{3}}
  \put(0,\qtopheight){\line(1,0){12}}
\ifcase #1 
    \put(6,0){\line(0,1){\qheight}}
\or \put(6,\qheight){\line(0,-1){6.5}}
\or \put(6,0){\line(0,1){6.5}}
\fi
\end{picture}
}

\newcommand{\ct}[1]{
\begin{picture}(12,\qheight)(0,0)
  \multiput(0,\qtopheight)(11,0){2}{\line(1,0){1}}
  \put(6,\qtopheight){\circle{\qheight}}
  \put(0,0){\vbox to \myvsize{\vfill
	\hbox to \myhsize{\hfill #1\hfill}\vfill}}
\end{picture}
}

\newcommand{\ti}[1]{
\begin{picture}(12,\qheight)(0,0)
  \multiput(1,0)(\qheight,0){2}{\line(0,1){\qheight}}
  \multiput(1,0)(0,\qheight){2}{\line(1,0){\qheight}}
  \multiput(0,\qtopheight)(11,0){2}{\line(1,0){1}}
  \put(0,0){\vbox to \myvsize{\vfill
	\hbox to \myhsize{\hfill #1\hfill}\vfill}}
\end{picture}
}

\newcommand{\tc}[1]{
\begin{picture}(12,\qheight)(0,0)
 \put(6,\qtopheight){\circle{\qheight}}
 \multiput(0,\qtopheight)(11,0){2}{\line(1,0){1}}
 \put(0,0){\vbox to \myvsize{\vfill
	\hbox to \myhsize{\hfill #1\hfill}\vfill}}
\end{picture}
}

\newcommand{\tb}[2]{
\begin{picture}(12,\qheight)(0,0)
  \put(1,0){\line(0,1){\qheight}}
  \qthicklines
  \put(11,0){\line(0,1){\qheight}}
  \thinlines
  \multiput(1,0)(\qheight,0){2}{\line(0,1){\qheight}}
  \multiput(0,\qtopheight)(11,0){2}{\line(1,0){1}}
  \put(0,0){\vbox to \myvsize{\vfill
	\hbox to \myhsize{\hfill #2\hfill}\vfill}}
\ifcase #1
{}
\or \put(1,0){\line(1,0){\qheight}}
\or \put(1,\qheight){\line(1,0){\qheight}}
\or \multiput(1,0)(0,\qheight){2}{\line(1,0){\qheight}}
\fi
\end{picture}
}

\newcommand{\tp}[2]{
\begin{picture}(12,\qheight)(0,0)
  \qthicklines
  \put(1,0){\line(0,1){\qheight}}
  \thinlines
  \put(11,0){\line(0,1){\qheight}}
  \multiput(0,\qtopheight)(11,0){2}{\line(1,0){1}}
  \put(0,0){\vbox to \myvsize{\vfill
	\hbox to \myhsize{\hfill #2\hfill}\vfill}}
\ifcase #1
{}
\or \put(1,0){\line(1,0){\qheight}}
\or \put(1,\qheight){\line(1,0){\qheight}}
\or \multiput(1,0)(0,\qheight){2}{\line(1,0){\qheight}}
\fi
\end{picture}
}

\newcommand{\place}[1]{\vbox to \myvsize{\vfill
	\hbox to \myhsize{\hfill #1\hfill}\vfill}}
\def\plac#1#2{\vbox to \myvsize{\vfill
	\hbox to #1\myhsize{#2\hfill}\vfill}}

\newcommand{\xor}{\mathbin{\oplus}}
\newcommand{\maj}{\mathop{\rm MAJ}\nolimits}
\newcommand{\uma}{\mathop{\rm UMA}\nolimits}
\newcommand{\xoreq}{\mathbin{\oplus\!\!=}}
\newcommand{\Z}{\mathbb Z}
 
\newcommand{\NOT}{{\sc NOT}}
\newcommand{\CNOT}{{\sc CNOT}}

\title{A new quantum ripple-carry addition circuit}
\author{Steven A.~Cuccaro\thanks{Center for Computing Sciences, 17100 Science Drive, Bowie, MD 20715.  {\tt cuccaro@super.org}}
\and Thomas G.~Draper\thanks{Department of Mathematics, University of Maryland, College Park, MD 20742.  {\tt tgd@math.umd.edu}}
\and Samuel A.~Kutin\thanks{Center for Communications Research, 805 Bunn Drive,
Princeton, NJ 08540. {\tt kutin@idaccr.org}}
\and David Petrie Moulton\thanks{Center for Communications Research, 805 Bunn Drive,
Princeton, NJ 08540. {\tt moulton@idaccr.org}}}
\begin{document}

\maketitle
\begin{abstract} 
We present a new linear-depth ripple-carry quantum addition circuit.
Previous addition circuits required linearly many ancillary qubits;
our new adder uses only a single ancillary qubit.  Also, our circuit has
lower depth and fewer gates than previous ripple-carry adders.
\end{abstract}

\section{Introduction}
\label{intro-sec}

We present a new quantum circuit for addition.  The circuit is
based on the {\em ripple-carry} approach, in which we start with the
low-order bits of the input and work our way up to the high-order
bits.  Since our computation must be reversible, we then work our way
from the high-order bits back down to the low-order bits.

A ripple-carry adder has previously been proposed by Vedral,
Barenco, and Ekert~\cite{VBE}.  Their circuit takes two $n$-bit
numbers as input, computes the sum in place, and outputs a single
bit (the high bit of the sum).  They also require $n - O(1)$
 scratch qubits, or {\em ancillae\/}.

Our circuit is different in that it requires only one ancilla.
Also, the depth and size of the circuit are smaller. The VBE adder is
made up of $4n + O(1)$ {\CNOT} (controlled-{\NOT}) gates and $4n + O(1)$
Toffoli (doubly-controlled-{\NOT}) gates,
with little parallelism.  Our circuit uses $2n + O(1)$ Toffoli
gates, $5n + O(1)$ {\CNOT} gates, and $2n+O(1)$ negations; the
depth is $2n + O(1)$.

The key ingredient of the new adder is a circuit computing the
majority of three bits in place.  We present this circuit, and a
simple version of the adder, in Section~\ref{basic-sec}.  We then
give an optimized version in Section~\ref{main-sec}.
See Figure~\ref{pseudocode-fig} for a pseudocode version of
the adder, and Figure~\ref{best-ripple-fig} for a pictorial version.

In Section~\ref{extensions-sec}, we discuss several variants of the
adder:\ performing addition modulo $2^n$, using an incoming carry bit,
and computing only the high bit of the sum.  This last variant can
be modified to produce a comparator.  The complexities of these
variants are summarized in Table~\ref{summary-table} on
Page~\pageref{summary-table}.

\section{The basic idea}
\label{basic-sec}

Our goal is to compute the sum of two $n$-bit numbers $a$ and $b$.
Write $a = a_{n-1} \cdots a_0$, with $a_0$ the lowest-order
bit, and similarly write $b = b_{n-1} \cdots b_0$.  We use
$A_i$ and $B_i$ to denote the memory locations where $a_i$ and $b_i$
are initially stored.

We will add $a$ and $b$ in place; at the end, $B_i$ will contain
$s_i$, the $i$th bit of the sum.  There is one additional output
location, $Z$, for the high bit $s_n$.

We define the {\em carry string\/} for $a$ and $b$ recursively:  Let
$c_0 = 0$, and let $c_{i+1} = \maj(a_i, b_i, c_i)$ for $i \ge 0$.
Note that $\maj(a_i, b_i, c_i) = a_i b_i \xor a_i c_i \xor b_i c_i$. 
We then have $s_i = a_i \xor b_i \xor c_i$ for all $i < n$, and $s_n = c_n$.
In a classical ripple-carry adder, we compute each $c_i$ in order,
working our way from $c_1$ up to $c_n$.  In a reversible ripple-carry
adder, we must then erase the carry bits, working our way back down.

\begin{figure}[h!]
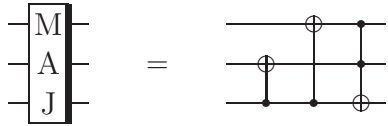

\begin{center}
\renewcommand{\arraystretch}{0}
\begin{tabular}{r@{}*{11}{c@{}}l}
 &\sn0&\tb2 M&\sn0& \wire{${}$} &\sm0&\sn0&\sm0&\xn2&\sm0&\cn2&\sn0& \\
 &\sn0&\tb0 A&\sn0& \wire{$\quad\quad{=}\quad\quad$} &\sm0&\xn2&\sm0&\sn2&\sm0&\cn0&\sn0& \\
 &\sn0&\tb1 J&\sn0& \wire{${}$} &\sm0&\cn1&\sm0&\cn1&\sm0&\xn1&\sn0& \\
\end{tabular}

\end{center}
\caption{The in-place majority gate ${\maj}$}
\label{maj3-fig}
\end{figure}

The first component of our adder, depicted in Figure~\ref{maj3-fig},
is a gate that computes the majority of three bits in place.  We build
our circuits out of negations, {\CNOT}s, and Toffoli
gates; time flows from left to right in our circuit diagrams.  For the
in-place majority, we apply first two {\CNOT}s and then one Toffoli.

\begin{figure}[h!]
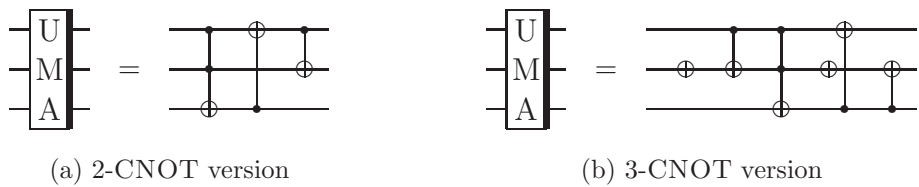

\begin{center}
\mbox{\subfigure[2-{\CNOT} version]{\renewcommand{\arraystretch}{0}
\begin{tabular}{r@{}*{11}{c@{}}l}
 &\sn0&\tb2 U&\sn0& \wire{${}$} &\sm0&\cn2&\sm0&\xn2&\sm0&\cn2&\sn0& \\
 &\sn0&\tb0 M&\sn0& \wire{$\quad{=}\quad$} &\sm0&\cn0&\sm0&\sn2&\sm0&\xn1&\sn0& \\
 &\sn0&\tb1 A&\sn0& \wire{${}$} &\sm0&\xn1&\sm0&\cn1&\sm0&\sn0&\sn0& \\
\end{tabular}
}\hspace{.5in}
\subfigure[3-{\CNOT} version]{\renewcommand{\arraystretch}{0}
\begin{tabular}{r@{}*{16}{c@{}}l}
 &\sn0&\tb2 U&\sn0& \wire{${}$} &\sm0&\sn0&\sm0&\cn2&\sm0&\cn2&\sm0&\sn0&\xn2&\sm0&\sn0&\sn0& \\
 &\sn0&\tb0 M&\sn0& \wire{$\quad{=}\quad$} &\sm0&\xn3&\sm0&\xn1&\sm0&\cn0&\sm0&\xn3&\sn2&\sm0&\xn2&\sn0& \\
 &\sn0&\tb1 A&\sn0& \wire{${}$} &\sm0&\sn0&\sm0&\sn0&\sm0&\xn1&\sm0&\sn0&\cn1&\sm0&\cn1&\sn0& \\
\end{tabular}
\label{fig-2b}}}
\end{center}
\caption{Two implementations of the ${\uma}$ gate}
\label{uma-fig}
\end{figure}

The second component, depicted in Figure~\ref{uma-fig}, is an
``UnMajority and Add'', or $\uma$, gate.  We give two versions, each
of which computes the same function on the qubits.  The first is
conceptually simpler, but the second admits greater parallelism.

\begin{figure}[h!]
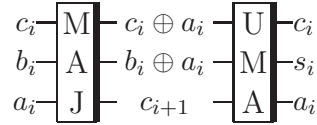

\begin{center}
\renewcommand{\arraystretch}{0}
\begin{tabular}{r@{}*{7}{c@{}}l}
\wire{$c_i$} &\sn0&\tb2 M&\sn0& \wire{$\;c_i\xor{a_i}\;$} &\sn0&\tb2 U&\sn0& \wire{$c_i$} \\
\wire{$b_i$} &\sn0&\tb0 A&\sn0& \wire{$b_i\xor{a_i}$} &\sn0&\tb0 M&\sn0& \wire{$s_i$} \\
\wire{$a_i$} &\sn0&\tb1 J&\sn0& \wire{$c_{i+1}$} &\sn0&\tb1 A&\sn0& \wire{$a_i$} \\
\end{tabular}

\end{center}
\caption{Combining the ${\maj}$ and ${\uma}$ gates}
\label{majuma-fig}
\end{figure}

The effect of using these two gates together is shown in
Figure~\ref{majuma-fig}.  Suppose that we have just computed the
carry bit $c_i$.  We apply the ${\maj}$ gate, which writes
$c_{i+1}$ into $A_i$.  We then continue our computation.
After we are done using $c_{i+1}$, we apply the ${\uma}$ gate,
which restores $a_i$ to $A_i$ and $c_i$ to $A_{i-1}$ and
writes $s_i$ to $B_i$.

\begin{figure}[h!]
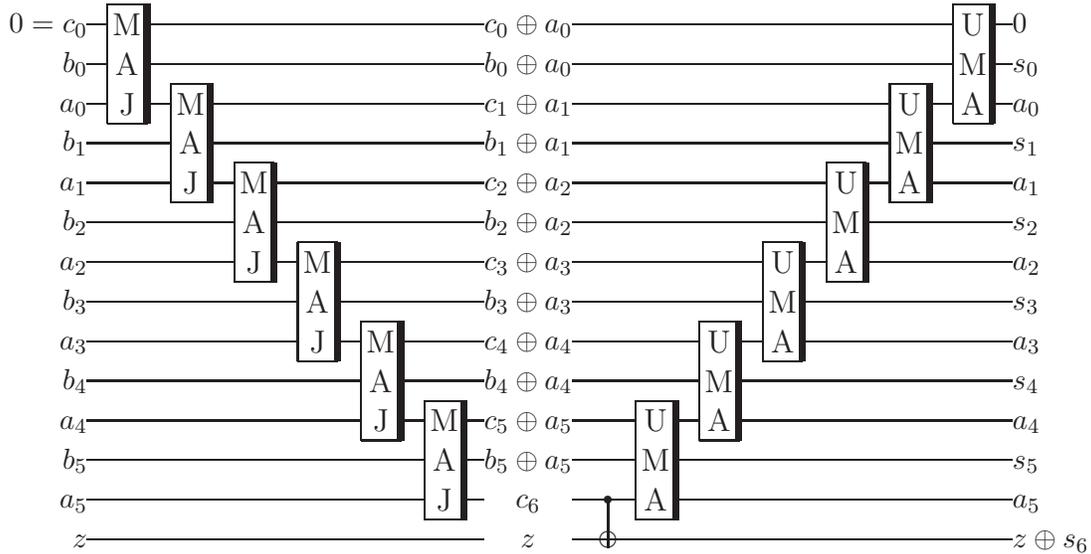

\begin{center}
\renewcommand{\arraystretch}{0}
\renewcommand{\smwidth}{7}
\renewcommand{\smhalf}{3}
\begin{tabular}{r@{}*{29}{c@{}}l}
\wire{$0=c_0$} &\sn0&\tb2 M&\sn0&\sx0&\sn0&\sx0&\sn0&\sx0&\sn0&\sx0&\sn0&\sx0&\sn0& \wire{$c_0\xor{a_0}$} &\sm0&\sn0&\sn0&\sx0&\sn0&\sx0&\sn0&\sx0&\sn0&\sx0&\sn0&\sx0&\sn0&\tb2 U&\sn0& \wire{$0$} \\
\wire{$b_0$} &\sn0&\tb0 A&\sn0&\sx0&\sn0&\sx0&\sn0&\sx0&\sn0&\sx0&\sn0&\sx0&\sn0& \wire{$b_0\xor{a_0}$} &\sm0&\sn0&\sn0&\sx0&\sn0&\sx0&\sn0&\sx0&\sn0&\sx0&\sn0&\sx0&\sn0&\tb0 M&\sn0& \wire{$s_0$} \\
\wire{$a_0$} &\sn0&\tb1 J&\sn0&\tb2 M&\sn0&\sx0&\sn0&\sx0&\sn0&\sx0&\sn0&\sx0&\sn0& \wire{$c_1\xor{a_1}$} &\sm0&\sn0&\sn0&\sx0&\sn0&\sx0&\sn0&\sx0&\sn0&\sx0&\sn0&\tb2 U&\sn0&\tb1 A&\sn0& \wire{$a_0$} \\
\wire{$b_1$} &\sn0&\sx0&\sn0&\tb0 A&\sn0&\sx0&\sn0&\sx0&\sn0&\sx0&\sn0&\sx0&\sn0& \wire{$b_1\xor{a_1}$} &\sm0&\sn0&\sn0&\sx0&\sn0&\sx0&\sn0&\sx0&\sn0&\sx0&\sn0&\tb0 M&\sn0&\sx0&\sn0& \wire{$s_1$} \\
\wire{$a_1$} &\sn0&\sx0&\sn0&\tb1 J&\sn0&\tb2 M&\sn0&\sx0&\sn0&\sx0&\sn0&\sx0&\sn0& \wire{$c_2\xor{a_2}$} &\sm0&\sn0&\sn0&\sx0&\sn0&\sx0&\sn0&\sx0&\sn0&\tb2 U&\sn0&\tb1 A&\sn0&\sx0&\sn0& \wire{$a_1$} \\
\wire{$b_2$} &\sn0&\sx0&\sn0&\sx0&\sn0&\tb0 A&\sn0&\sx0&\sn0&\sx0&\sn0&\sx0&\sn0& \wire{$b_2\xor{a_2}$} &\sm0&\sn0&\sn0&\sx0&\sn0&\sx0&\sn0&\sx0&\sn0&\tb0 M&\sn0&\sx0&\sn0&\sx0&\sn0& \wire{$s_2$} \\
\wire{$a_2$} &\sn0&\sx0&\sn0&\sx0&\sn0&\tb1 J&\sn0&\tb2 M&\sn0&\sx0&\sn0&\sx0&\sn0& \wire{$c_3\xor{a_3}$} &\sm0&\sn0&\sn0&\sx0&\sn0&\sx0&\sn0&\tb2 U&\sn0&\tb1 A&\sn0&\sx0&\sn0&\sx0&\sn0& \wire{$a_2$} \\
\wire{$b_3$} &\sn0&\sx0&\sn0&\sx0&\sn0&\sx0&\sn0&\tb0 A&\sn0&\sx0&\sn0&\sx0&\sn0& \wire{$b_3\xor{a_3}$} &\sm0&\sn0&\sn0&\sx0&\sn0&\sx0&\sn0&\tb0 M&\sn0&\sx0&\sn0&\sx0&\sn0&\sx0&\sn0& \wire{$s_3$} \\
\wire{$a_3$} &\sn0&\sx0&\sn0&\sx0&\sn0&\sx0&\sn0&\tb1 J&\sn0&\tb2 M&\sn0&\sx0&\sn0& \wire{$c_4\xor{a_4}$} &\sm0&\sn0&\sn0&\sx0&\sn0&\tb2 U&\sn0&\tb1 A&\sn0&\sx0&\sn0&\sx0&\sn0&\sx0&\sn0& \wire{$a_3$} \\
\wire{$b_4$} &\sn0&\sx0&\sn0&\sx0&\sn0&\sx0&\sn0&\sx0&\sn0&\tb0 A&\sn0&\sx0&\sn0& \wire{$b_4\xor{a_4}$} &\sm0&\sn0&\sn0&\sx0&\sn0&\tb0 M&\sn0&\sx0&\sn0&\sx0&\sn0&\sx0&\sn0&\sx0&\sn0& \wire{$s_4$} \\
\wire{$a_4$} &\sn0&\sx0&\sn0&\sx0&\sn0&\sx0&\sn0&\sx0&\sn0&\tb1 J&\sn0&\tb2 M&\sn0& \wire{$c_5\xor{a_5}$} &\sm0&\sn0&\sn0&\tb2 U&\sn0&\tb1 A&\sn0&\sx0&\sn0&\sx0&\sn0&\sx0&\sn0&\sx0&\sn0& \wire{$a_4$} \\
\wire{$b_5$} &\sn0&\sx0&\sn0&\sx0&\sn0&\sx0&\sn0&\sx0&\sn0&\sx0&\sn0&\tb0 A&\sn0& \wire{$b_5\xor{a_5}$} &\sm0&\sn0&\sn0&\tb0 M&\sn0&\sx0&\sn0&\sx0&\sn0&\sx0&\sn0&\sx0&\sn0&\sx0&\sn0& \wire{$s_5$} \\
\wire{$a_5$} &\sn0&\sx0&\sn0&\sx0&\sn0&\sx0&\sn0&\sx0&\sn0&\sx0&\sn0&\tb1 J&\sn0& \wire{$c_6$} &\sm0&\cn2&\sn0&\tb1 A&\sn0&\sx0&\sn0&\sx0&\sn0&\sx0&\sn0&\sx0&\sn0&\sx0&\sn0& \wire{$a_5$} \\
\wire{$z$} &\sn0&\sx0&\sn0&\sx0&\sn0&\sx0&\sn0&\sx0&\sn0&\sx0&\sn0&\sx0&\sn0& \wire{$z$} &\sm0&\xn1&\sn0&\sx0&\sn0&\sx0&\sn0&\sx0&\sn0&\sx0&\sn0&\sx0&\sn0&\sx0&\sn0& \wire{$z\xor{s_6}$} \\
\end{tabular}

\end{center}
\caption{A simple ripple-carry adder for $n = 6$.}
\label{simple-ripple-fig}
\end{figure}

It follows that we can string together ${\maj}$ and ${\uma}$ gates
to build a ripple-carry adder.  Such an adder is depicted in
Figure~\ref{simple-ripple-fig}.  We have one ancilla, labeled $X$,
initialized to $0$.  We view $X$ as containing the initial carry bit $c_0$.
The output bit $Z$ contains some value $z$ when the circuit begins and
$z \xor s_n$ when the circuit concludes.

\section{Improving the circuit}
\label{main-sec}

We can reduce the depth of the basic circuit of Figure~\ref{simple-ripple-fig}
in several ways.  It is necessary to use the 3-{\CNOT} version of the
${\uma}$ gate from Figure~\ref{fig-2b}.

\newcommand{\csep}{\quad;\quad}
\begin{figure}[p]
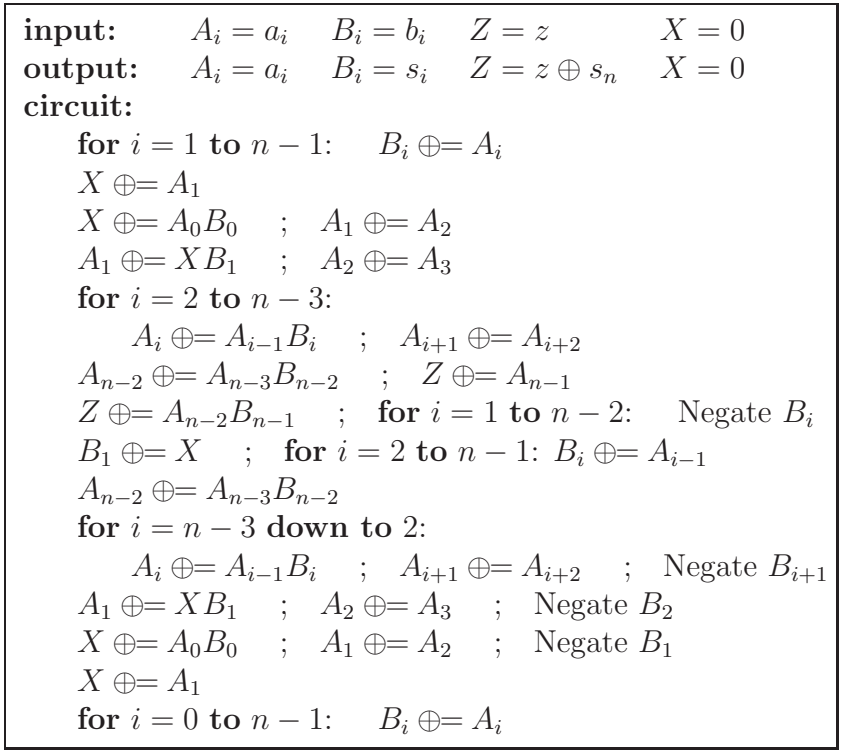

\centerline{\fbox{
\begin{minipage}{100in}
\renewcommand{\baselinestretch}{1.5}
\begin{tabbing}
\bf{output:} \quad \= $A_i = a_i$ \quad \= $B_i = b_i$ \quad \= $Z = z \xor s_n$ \quad \= $X = 0$ \quad \kill
\bf{input:} \> $A_i = a_i$ \> $B_i = b_i$ \> $Z = z$ \> $X = 0$ \\
\bf{output:} \> $A_i = a_i$ \> $B_i = s_i$ \> $Z = z \xor s_n$ \> $X = 0$ \\
\bf{circuit:} \\
{\bf for} \=\+\kill
{\bf for} \= $i = 1$ {\bf to} $n-1$: \quad $B_i \xoreq A_i$ \\
$X \xoreq A_1$ \\
$X \xoreq A_0 B_0$ \csep $A_1 \xoreq A_2$ \\
$A_1 \xoreq X B_1$ \csep $A_2 \xoreq A_3$ \\
{\bf for} $i = 2$ {\bf to} $n-3$: \\
\> $A_i \xoreq A_{i-1} B_i$ \csep $A_{i+1} \xoreq A_{i+2}$ \\
$A_{n-2} \xoreq A_{n-3} B_{n-2}$ \csep $Z \xoreq A_{n-1}$ \\
$Z \xoreq A_{n-2} B_{n-1}$ \csep
{\bf for} \= $i = 1$ {\bf to} $n-2$: \quad Negate $B_i$ \\
$B_1 \xoreq X$ \csep
{\bf for} \= $i = 2$ {\bf to} $n-1$: $B_i \xoreq A_{i-1}$ \\
$A_{n-2} \xoreq A_{n-3} B_{n-2}$ \\
{\bf for} \= $i = n-3$ {\bf down to} $2$: \\
\> $A_i \xoreq A_{i-1} B_i$ \csep $A_{i+1} \xoreq A_{i+2}$
\csep Negate $B_{i+1}$ \\
$A_1 \xoreq X B_1$ \csep $A_2 \xoreq A_3$ \csep Negate $B_2$ \\
$X \xoreq A_0 B_0$ \csep $A_1 \xoreq A_2$ \csep Negate $B_1$ \\
$X \xoreq A_1$ \\
{\bf for} $i = 0$ {\bf to} $n-1$: \quad $B_i \xoreq A_i$
\end{tabbing}
\end{minipage}}}
\caption{The ripple-carry adder for $n \ge 4$.  Each line of pseudocode
corresponds to a single time-slice.}
\label{pseudocode-fig}
\end{figure}

\begin{figure}[p]
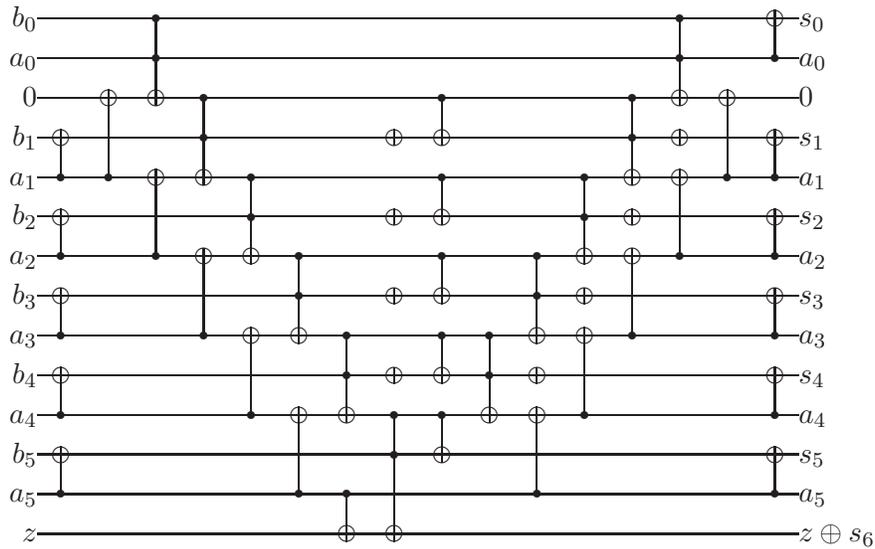

\begin{center}
\renewcommand{\arraystretch}{0}
\begin{tabular}{r@{}*{33}{c@{}}l}
\wire{$b_0$} &\sn0&\sn0&\sm0&\sn0&\sm0&\cn2&\sm0&\sn0&\sm0&\sn0&\sm0&\sn0&\sm0&\sn0&\sm0&\sn0&\sm0&\sn0&\sm0&\sn0&\sm0&\sn0&\sm0&\sn0&\sm0&\sn0&\sm0&\cn2&\sm0&\sn0&\sm0&\xn2&\sn0& \wire{$s_0$} \\
\wire{$a_0$} &\sn0&\sn0&\sm0&\sn0&\sm0&\cn0&\sm0&\sn0&\sm0&\sn0&\sm0&\sn0&\sm0&\sn0&\sm0&\sn0&\sm0&\sn0&\sm0&\sn0&\sm0&\sn0&\sm0&\sn0&\sm0&\sn0&\sm0&\cn0&\sm0&\sn0&\sm0&\cn1&\sn0& \wire{$a_0$} \\
\wire{$0$} &\sn0&\sn0&\sm0&\xn2&\sm0&\xn1&\sm0&\cn2&\sm0&\sn0&\sm0&\sn0&\sm0&\sn0&\sm0&\sn0&\sm0&\cn2&\sm0&\sn0&\sm0&\sn0&\sm0&\sn0&\sm0&\cn2&\sm0&\xn1&\sm0&\xn2&\sm0&\sn0&\sn0& \wire{$0$} \\
\wire{$b_1$} &\sn0&\xn2&\sm0&\sn2&\sm0&\sn0&\sm0&\cn0&\sm0&\sn0&\sm0&\sn0&\sm0&\sn0&\sm0&\xn3&\sm0&\xn1&\sm0&\sn0&\sm0&\sn0&\sm0&\sn0&\sm0&\cn0&\sm0&\xn3&\sm0&\sn2&\sm0&\xn2&\sn0& \wire{$s_1$} \\
\wire{$a_1$} &\sn0&\cn1&\sm0&\cn1&\sm0&\xn2&\sm0&\xn1&\sm0&\cn2&\sm0&\sn0&\sm0&\sn0&\sm0&\sn0&\sm0&\cn2&\sm0&\sn0&\sm0&\sn0&\sm0&\cn2&\sm0&\xn1&\sm0&\xn2&\sm0&\cn1&\sm0&\cn1&\sn0& \wire{$a_1$} \\
\wire{$b_2$} &\sn0&\xn2&\sm0&\sn0&\sm0&\sn2&\sm0&\sn0&\sm0&\cn0&\sm0&\sn0&\sm0&\sn0&\sm0&\xn3&\sm0&\xn1&\sm0&\sn0&\sm0&\sn0&\sm0&\cn0&\sm0&\xn3&\sm0&\sn2&\sm0&\sn0&\sm0&\xn2&\sn0& \wire{$s_2$} \\
\wire{$a_2$} &\sn0&\cn1&\sm0&\sn0&\sm0&\cn1&\sm0&\xn2&\sm0&\xn1&\sm0&\cn2&\sm0&\sn0&\sm0&\sn0&\sm0&\cn2&\sm0&\sn0&\sm0&\cn2&\sm0&\xn1&\sm0&\xn2&\sm0&\cn1&\sm0&\sn0&\sm0&\cn1&\sn0& \wire{$a_2$} \\
\wire{$b_3$} &\sn0&\xn2&\sm0&\sn0&\sm0&\sn0&\sm0&\sn2&\sm0&\sn0&\sm0&\cn0&\sm0&\sn0&\sm0&\xn3&\sm0&\xn1&\sm0&\sn0&\sm0&\cn0&\sm0&\xn3&\sm0&\sn2&\sm0&\sn0&\sm0&\sn0&\sm0&\xn2&\sn0& \wire{$s_3$} \\
\wire{$a_3$} &\sn0&\cn1&\sm0&\sn0&\sm0&\sn0&\sm0&\cn1&\sm0&\xn2&\sm0&\xn1&\sm0&\cn2&\sm0&\sn0&\sm0&\cn2&\sm0&\cn2&\sm0&\xn1&\sm0&\xn2&\sm0&\cn1&\sm0&\sn0&\sm0&\sn0&\sm0&\cn1&\sn0& \wire{$a_3$} \\
\wire{$b_4$} &\sn0&\xn2&\sm0&\sn0&\sm0&\sn0&\sm0&\sn0&\sm0&\sn2&\sm0&\sn0&\sm0&\cn0&\sm0&\xn3&\sm0&\xn1&\sm0&\cn0&\sm0&\xn3&\sm0&\sn2&\sm0&\sn0&\sm0&\sn0&\sm0&\sn0&\sm0&\xn2&\sn0& \wire{$s_4$} \\
\wire{$a_4$} &\sn0&\cn1&\sm0&\sn0&\sm0&\sn0&\sm0&\sn0&\sm0&\cn1&\sm0&\xn2&\sm0&\xn1&\sm0&\cn2&\sm0&\cn2&\sm0&\xn1&\sm0&\xn2&\sm0&\cn1&\sm0&\sn0&\sm0&\sn0&\sm0&\sn0&\sm0&\cn1&\sn0& \wire{$a_4$} \\
\wire{$b_5$} &\sn0&\xn2&\sm0&\sn0&\sm0&\sn0&\sm0&\sn0&\sm0&\sn0&\sm0&\sn2&\sm0&\sn0&\sm0&\cn0&\sm0&\xn1&\sm0&\sn0&\sm0&\sn2&\sm0&\sn0&\sm0&\sn0&\sm0&\sn0&\sm0&\sn0&\sm0&\xn2&\sn0& \wire{$s_5$} \\
\wire{$a_5$} &\sn0&\cn1&\sm0&\sn0&\sm0&\sn0&\sm0&\sn0&\sm0&\sn0&\sm0&\cn1&\sm0&\cn2&\sm0&\sn2&\sm0&\sn0&\sm0&\sn0&\sm0&\cn1&\sm0&\sn0&\sm0&\sn0&\sm0&\sn0&\sm0&\sn0&\sm0&\cn1&\sn0& \wire{$a_5$} \\
\wire{$z$} &\sn0&\sn0&\sm0&\sn0&\sm0&\sn0&\sm0&\sn0&\sm0&\sn0&\sm0&\sn0&\sm0&\xn1&\sm0&\xn1&\sm0&\sn0&\sm0&\sn0&\sm0&\sn0&\sm0&\sn0&\sm0&\sn0&\sm0&\sn0&\sm0&\sn0&\sm0&\sn0&\sn0& \wire{$z\xor{s_6}$} \\
\end{tabular}

\end{center}
\caption{The ripple-carry adder for $n = 6$.}
\label{best-ripple-fig}
\end{figure}

\begin{enumerate}
\item The first {\CNOT}s of all the ${\maj}$ gates can be performed
in a single time-slice at the beginning.  Similarly, the final
{\CNOT}s of all the ${\uma}$ gates can be performed in a single
time-slice at the end.
\item Consider the first half of the circuit:\ the ${\maj}$ ripple.
The Toffoli at the end of the $i$th ${\maj}$ gate
commutes with the second {\CNOT} of the $(i+1)$th gate.  If we swap these
two gates for each $i$, then the depth decreases:\ the Toffoli of
the $i$th ${\maj}$ gate can now be done in parallel
with the second {\CNOT} of the $(i+2)$th ${\maj}$ gate.
\item We can perform a similar transformation
on the second half of the circuit.
We swap the Toffoli of the $(i+1)$th ${\uma}$ gate with the second
{\CNOT} of the $i$th ${\uma}$ gate.  Again, the depth decreases:\
the second {\CNOT} of the $i$th ${\uma}$ gate can be done in parallel
with the Toffoli of the $(i+2)$th ${\uma}$ gate.
\item We know $c_0 = 0$, so we do not need a $\maj$ gate to compute
$c_1 = a_0 b_0$.  Instead, we compute $c_1$ with a single Toffoli and
store it in our ancilla.  At the end of the circuit, we undo this
same Toffoli, and then set $B_0$ to $s_0$ with a single {\CNOT}.
\item It is inefficient to write $c_n$ into $A_{n-1}$, copy it to the
output, and then erase it.  We
can instead write directly to the output.  We replace the central
piece (two Toffolis, two {\CNOT}s, and two negations) with one
Toffoli and two {\CNOT}s.  One of the {\CNOT}s can be done in parallel
with other computation.
\end{enumerate}

Our final ripple-carry circuit is described in
Figure~\ref{pseudocode-fig}.  The construction applies for any $n$,
but the pseudocode in Figure~\ref{pseudocode-fig} is valid only for
$n\geq 4$.  A sample circuit for $n=6$ is depicted in
Figure~\ref{best-ripple-fig}.  Note that, in Figure~\ref{simple-ripple-fig},
the ancilla contains $c_0$ and is the topmost wire; in
Figure~\ref{best-ripple-fig}, the ancilla contains $c_1$ and is the
third wire from the top.

Assuming $n \ge 2$, the circuit size is $2n-1$ Toffoli gates, $5n
- 3$ {\CNOT}s, and $2n-4$ negations.  The depth is $2n+4$:\ $2n-1$ Toffoli
time-slices and $5$ {\CNOT} time-slices.

\suppressfloats

\section{Extensions}
\label{extensions-sec}

We now discuss various slightly-modified versions of the ripple-carry
adder:
\begin{itemize}
\item modulo $2^n$:  We do not compute the high bit.
\item incoming carry:  We consider the ancilla $c_0$ to be an
extra input bit.
\item high bit only:  We compute the high bit, but do not overwrite the
$b$ input.  This circuit can be adapted to give a comparator.
\end{itemize}

\begin{table}[t]
\begin{center}
\renewcommand{\arraystretch}{1.5}
\begin{tabular}{|c|c||ccc||c|c||c|}\hline
& & \multicolumn{3}{c||}{Number of Bits} &
\multicolumn{2}{c||}{Size} & Depth \\
Function & IC? & In & Out & Anc. & Toffoli & {\CNOT} & \\ \hline \hline
$+$ in $\Z$ & N & $2n$ & 1 & 1 & $2n-1$ & $5n-3$ & $2n+4$ \\
$+$ in $\Z$ & Y & $2n+1$ & 1 & 0 & $2n-1$ & $5n+1$ & $2n+6$ \\ \hline
$+$ (mod $2^n$) & N & $2n$ & 0 & 1 & $2n-3$ & $5n-7$ & $2n+2$ \\
$+$ (mod $2^n$) & Y & $2n+1$ & 0 & 0 & $2n-3$ & $5n-3$ & $2n+4$ \\ \hline
Compare & N & $2n$ & 1 & 1 & $2n-1$ & $4n-3$ & $2n+3$ \\
Compare & Y & $2n+1$ & 1 & 0 & $2n-1$ & $4n+1$ & $2n+5$ \\ \hline
VBE adder~\cite{VBE} & N & $2n$ & 1 & $n$ & $4n-2$ & $4n-2$ & $6n-2$ \\ \hline
\end{tabular}
\end{center}
\caption{Circuit summary, for $n \ge 3$.  The first column gives the function
being computed.  The second lists whether we take an incoming carry bit
as input.  We then list the number of input, output, and ancilla bits,
the number of Toffoli and {\CNOT} gates, and the overall depth.  We do
not include negations when counting size or depth. }
\label{summary-table}
\end{table}

In each case, the circuit is a simple modification of the circuit of
Section~\ref{main-sec}.  The only question is the exact
depth and size of the circuit.  The results are summarized in
Table~\ref{summary-table}.  For each circuit, we give the number of
Toffoli gates, the number of {\CNOT} gates, and the overall depth.  In
each case, the number of Toffoli time-slices is equal to the number
of Toffoli gates; the remaining time-slices contain {\CNOT}s.
For the VBE adder, the circuit has $3n - 1$ Toffoli time-slices and
$3n - 1$ {\CNOT} time-slices.

\subsection{Addition Modulo $2^n$}
\label{mod-2n-sec}

Suppose that we wish to compute $a+b\ (\text{mod } 2^n)$; that is, we do not
want to compute the high bit $c_n$.  One approach is the following:
\begin{enumerate}
\item \label{first-2n-step}Add the low-order $n-1$ bits of
$a$ and $b$, using the circuit of Section~\ref{main-sec}.
Use $B_{n-1}$ as the output bit.
\item \label{second-2n-step}Set $B_{n-1} \xoreq A_{n-1}$.
\end{enumerate}
After step 1, we have correctly computed $s_0$ through $s_{n-2}$, and
we have written $b_{n-1} \xor c_{n-1}$ into $B_{n-1}$.  Then,
in step~\ref{second-2n-step}, we complete the calculation of $s_{n-1}$.
Note that step~\ref{second-2n-step} occurs in parallel with the final
time-slice of step~\ref{first-2n-step}.

For $n \ge 3$, this circuit contains $2n - 3$ Toffolis,
$5n - 7$ {\CNOT}s, and $2n - 6$ negations.
The depth is $2n+2$:\  $2n - 3$ Toffoli
time-slices and 5 {\CNOT} time-slices.

\subsection{Addition with Incoming Carry}
\label{incoming-carry-sec}

Suppose we want to allow an incoming carry into our addition circuit.
We have an additional input bit $y$, and we compute $a + b + y$.

We observe that the circuit of Section~\ref{basic-sec} already
solves this problem; we use $y$ in place of the ancilla $c_0$.
We then correctly compute $c_1$, and the ripple continues.

We cannot use the fourth improvement from Section~\ref{main-sec},
since we can no longer assume the incoming bit is zero.  The
other improvements still apply.

We obtain a ripple-carry adder with incoming carry which consists of
$2n - 1$ Toffolis, $5n+1$ {\CNOT}s, and $2n-2$ negations.
For $n \ge 2$, the circuit has depth $2n+6$:\
$2n - 1$ Toffoli time-slices and 7 {\CNOT}
time-slices.

We can also apply the incoming-carry modification to the circuit
of Section~\ref{mod-2n-sec}.  For $n \ge 3$, we get a circuit
with $2n - 3$ Toffolis, $5n - 3$ {\CNOT}s, and $2n - 4$ negations.  The depth is $2n+4$:\
$2n - 3$ Toffoli time-slices and 7 {\CNOT} time-slices.

\subsection{High Bit Only}
\label{comparison-sec}

We now consider the problem of computing only the high bit of the sum
$a + b$.  The first half of the circuit is identical to the first half
of our adder from Section~\ref{main-sec}:\ when we get to the
middle point, we have written the high bit to $Z$.  Now, we simply
undo the first half of the circuit.  We can view this as applying a
series of $\maj$ gates, followed by a Toffoli and a series of
$\maj^{-1}$ gates.

For $n \ge 2$, the resulting circuit contains $2n - 1$ Toffoli gates
and $4n - 3$ {\CNOT}s.  The depth is $2n+3$:\ $2n - 1$ Toffoli time-slices and
$4$ {\CNOT} time-slices.

We can combine the high-bit circuit with the incoming-carry modification
discussed in Section~\ref{incoming-carry-sec}.  We obtain a circuit
with $2n - 1$ Toffolis and $4n+1$ {\CNOT}s.  For $n \ge 2$, the depth
is $2n+5$:\  $2n - 1$ Toffoli time-slices and $6$ {\CNOT} time-slices.

It is worth noting that our ripple-carry adder can easily be turned
into a subtractor.  Whether we use one's-complement or two's-complement
arithmetic, we have the identity
$$
a - b = (a' + b)',
$$
where $'$ denotes bitwise complementation.  Hence, we can subtract by
adding two time-slices:\ complement $a$ at the start, and complement
$a$ and $s$ at the end.

If we combine this subtraction idea with the high-bit computer of
this section, we obtain a comparator:\ we compute the high bit of
$a - b$, which is 1 if and only if $a < b$.

\section{Conclusions}\label{conclude-sec}

One interesting open problem is to construct an optimal addition circuit.
In particular, if a reversible addition circuit uses just one ancilla,
must it have linear depth?  A logarithmic-depth adder has been
constructed using $2n$ ancillae~\cite{log-adder}; more generally, for
any $k > 0$, we can construct a family of circuits using $n/k$ ancillae
with depth $O(k + \log n)$.  Is there a
logarithmic-depth addition circuit family using only a constant number
of ancillae?  If not, can we prove a lower bound on depth?

A version of our ripple-carry adder has been proposed that uses no
ancillae~\cite{log-compare}.  That circuit requires that the output bit
be initialized to zero.  We do not know whether we can add in linear
depth with no ancillae and without this restriction on the output bit.

It would be interesting to compare the ripple-carry adder of this
paper to the transform adder~\cite{draper-add}.  Both circuits have
linear depth.  It is unclear which adder would be easier to implement
in practice; the answer depends on the relative costs of Toffoli gates
and controlled rotations.

\begin{sidewaysfigure}
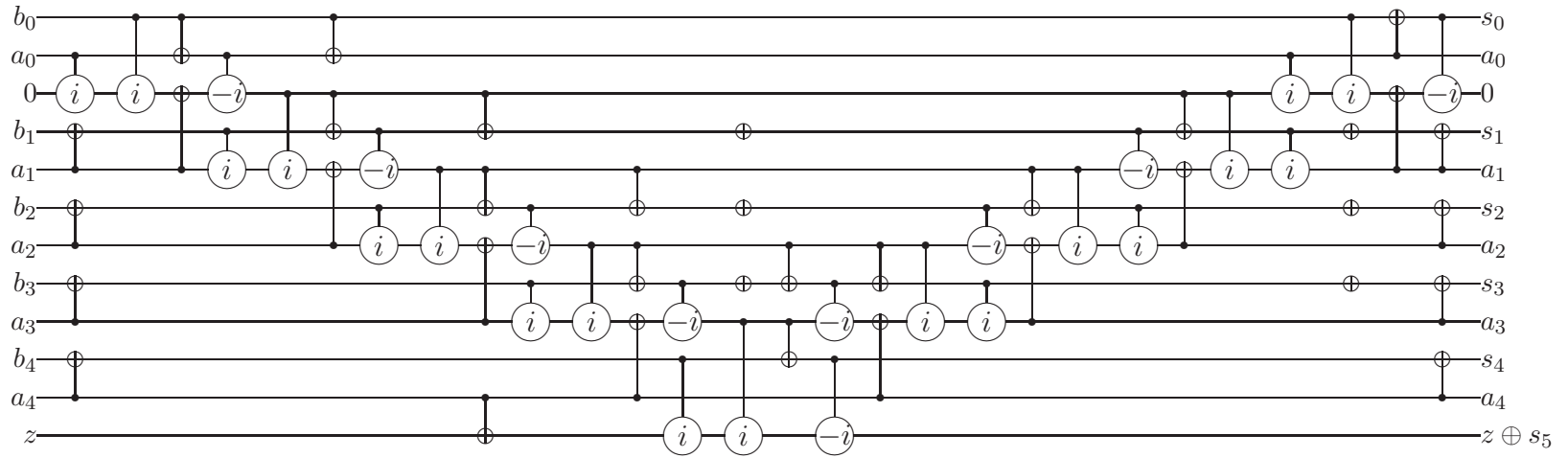

\begin{center}
\renewcommand{\arraystretch}{0}
\begin{tabular}{r@{}*{57}{c@{}}l}
\wire{$b_0$} &\sn0&\sx0&\sn0&\dx2&\sn0&\cn2&\sn0&\sx0&\sn0&\sx0&\sn0&\cn2&\sn0&\sx0&\sn0&\sx0&\sn0&\sn0&\sn0&\sx0&\sn0&\sx0&\sn0&\sn0&\sn0&\sx0&\sn0&\sx0&\sn0&\sn0&\sn0&\sx0&\sn0&\sn0&\sn0&\sx0&\sn0&\sx0&\sn0&\sn0&\sn0&\sx0&\sn0&\sx0&\sn0&\sn0&\sn0&\sx0&\sn0&\sx0&\sn0&\dx2&\sn0&\xn2&\sn0&\dx2&\sn0& \wire{$s_0$} \\
\wire{$a_0$} &\sn0&\dx2&\sn0&\sx2&\sn0&\xn1&\sn0&\dx2&\sn0&\sx0&\sn0&\xn1&\sn0&\sx0&\sn0&\sx0&\sn0&\sn0&\sn0&\sx0&\sn0&\sx0&\sn0&\sn0&\sn0&\sx0&\sn0&\sx0&\sn0&\sn0&\sn0&\sx0&\sn0&\sn0&\sn0&\sx0&\sn0&\sx0&\sn0&\sn0&\sn0&\sx0&\sn0&\sx0&\sn0&\sn0&\sn0&\sx0&\sn0&\dx2&\sn0&\sx2&\sn0&\cn1&\sn0&\sx2&\sn0& \wire{$a_0$} \\
\wire{$0$} &\sn0&\tc{$i$}&\sn0&\tc{$i$}&\sn0&\xn2&\sn0&\tc{$-i$}&\sn0&\dx2&\sn0&\cn2&\sn0&\sx0&\sn0&\sx0&\sn0&\cn2&\sn0&\sx0&\sn0&\sx0&\sn0&\sn0&\sn0&\sx0&\sn0&\sx0&\sn0&\sn0&\sn0&\sx0&\sn0&\sn0&\sn0&\sx0&\sn0&\sx0&\sn0&\sn0&\sn0&\sx0&\sn0&\sx0&\sn0&\cn2&\sn0&\dx2&\sn0&\tc{$i$}&\sn0&\tc{$i$}&\sn0&\xn2&\sn0&\tc{$-i$}&\sn0& \wire{$0$} \\
\wire{$b_1$} &\sn0&\nt2&\sn0&\sx0&\sn0&\sn2&\sn0&\dx2&\sn0&\sx2&\sn0&\xn1&\sn0&\dx2&\sn0&\sx0&\sn0&\xn1&\sn0&\sx0&\sn0&\sx0&\sn0&\sn0&\sn0&\sx0&\sn0&\nt3&\sn0&\sn0&\sn0&\sx0&\sn0&\sn0&\sn0&\sx0&\sn0&\sx0&\sn0&\sn0&\sn0&\sx0&\sn0&\dx2&\sn0&\xn1&\sn0&\sx2&\sn0&\dx2&\sn0&\nt3&\sn0&\sn2&\sn0&\nt2&\sn0& \wire{$s_1$} \\
\wire{$a_1$} &\sn0&\dx1&\sn0&\sx0&\sn0&\cn1&\sn0&\tc{$i$}&\sn0&\tc{$i$}&\sn0&\xn2&\sn0&\tc{$-i$}&\sn0&\dx2&\sn0&\cn2&\sn0&\sx0&\sn0&\sx0&\sn0&\cn2&\sn0&\sx0&\sn0&\sx0&\sn0&\sn0&\sn0&\sx0&\sn0&\sn0&\sn0&\sx0&\sn0&\sx0&\sn0&\cn2&\sn0&\dx2&\sn0&\tc{$-i$}&\sn0&\xn2&\sn0&\tc{$i$}&\sn0&\tc{$i$}&\sn0&\sx0&\sn0&\cn1&\sn0&\dx1&\sn0& \wire{$a_1$} \\
\wire{$b_2$} &\sn0&\nt2&\sn0&\sx0&\sn0&\sn0&\sn0&\sx0&\sn0&\sx0&\sn0&\sn2&\sn0&\dx2&\sn0&\sx2&\sn0&\xn1&\sn0&\dx2&\sn0&\sx0&\sn0&\xn1&\sn0&\sx0&\sn0&\nt3&\sn0&\sn0&\sn0&\sx0&\sn0&\sn0&\sn0&\sx0&\sn0&\dx2&\sn0&\xn1&\sn0&\sx2&\sn0&\dx2&\sn0&\sn2&\sn0&\sx0&\sn0&\sx0&\sn0&\nt3&\sn0&\sn0&\sn0&\nt2&\sn0& \wire{$s_2$} \\
\wire{$a_2$} &\sn0&\dx1&\sn0&\sx0&\sn0&\sn0&\sn0&\sx0&\sn0&\sx0&\sn0&\cn1&\sn0&\tc{$i$}&\sn0&\tc{$i$}&\sn0&\xn2&\sn0&\tc{$-i$}&\sn0&\dx2&\sn0&\cn2&\sn0&\sx0&\sn0&\sx0&\sn0&\cn2&\sn0&\sx0&\sn0&\cn2&\sn0&\dx2&\sn0&\tc{$-i$}&\sn0&\xn2&\sn0&\tc{$i$}&\sn0&\tc{$i$}&\sn0&\cn1&\sn0&\sx0&\sn0&\sx0&\sn0&\sx0&\sn0&\sn0&\sn0&\dx1&\sn0& \wire{$a_2$} \\
\wire{$b_3$} &\sn0&\nt2&\sn0&\sx0&\sn0&\sn0&\sn0&\sx0&\sn0&\sx0&\sn0&\sn0&\sn0&\sx0&\sn0&\sx0&\sn0&\sn2&\sn0&\dx2&\sn0&\sx2&\sn0&\xn1&\sn0&\dx2&\sn0&\nt3&\sn0&\xn1&\sn0&\dx2&\sn0&\xn1&\sn0&\sx2&\sn0&\dx2&\sn0&\sn2&\sn0&\sx0&\sn0&\sx0&\sn0&\sn0&\sn0&\sx0&\sn0&\sx0&\sn0&\nt3&\sn0&\sn0&\sn0&\nt2&\sn0& \wire{$s_3$} \\
\wire{$a_3$} &\sn0&\dx1&\sn0&\sx0&\sn0&\sn0&\sn0&\sx0&\sn0&\sx0&\sn0&\sn0&\sn0&\sx0&\sn0&\sx0&\sn0&\cn1&\sn0&\tc{$i$}&\sn0&\tc{$i$}&\sn0&\xn2&\sn0&\tc{$-i$}&\sn0&\dx2&\sn0&\cn2&\sn0&\tc{$-i$}&\sn0&\xn2&\sn0&\tc{$i$}&\sn0&\tc{$i$}&\sn0&\cn1&\sn0&\sx0&\sn0&\sx0&\sn0&\sn0&\sn0&\sx0&\sn0&\sx0&\sn0&\sx0&\sn0&\sn0&\sn0&\dx1&\sn0& \wire{$a_3$} \\
\wire{$b_4$} &\sn0&\nt2&\sn0&\sx0&\sn0&\sn0&\sn0&\sx0&\sn0&\sx0&\sn0&\sn0&\sn0&\sx0&\sn0&\sx0&\sn0&\sn0&\sn0&\sx0&\sn0&\sx0&\sn0&\sn2&\sn0&\dx2&\sn0&\sx2&\sn0&\xn1&\sn0&\dx2&\sn0&\sn2&\sn0&\sx0&\sn0&\sx0&\sn0&\sn0&\sn0&\sx0&\sn0&\sx0&\sn0&\sn0&\sn0&\sx0&\sn0&\sx0&\sn0&\sx0&\sn0&\sn0&\sn0&\nt2&\sn0& \wire{$s_4$} \\
\wire{$a_4$} &\sn0&\dx1&\sn0&\sx0&\sn0&\sn0&\sn0&\sx0&\sn0&\sx0&\sn0&\sn0&\sn0&\sx0&\sn0&\sx0&\sn0&\cn2&\sn0&\sx0&\sn0&\sx0&\sn0&\cn1&\sn0&\sx2&\sn0&\sx2&\sn0&\sn0&\sn0&\sx2&\sn0&\cn1&\sn0&\sx0&\sn0&\sx0&\sn0&\sn0&\sn0&\sx0&\sn0&\sx0&\sn0&\sn0&\sn0&\sx0&\sn0&\sx0&\sn0&\sx0&\sn0&\sn0&\sn0&\dx1&\sn0& \wire{$a_4$} \\
\wire{$z$} &\sn0&\sx0&\sn0&\sx0&\sn0&\sn0&\sn0&\sx0&\sn0&\sx0&\sn0&\sn0&\sn0&\sx0&\sn0&\sx0&\sn0&\xn1&\sn0&\sx0&\sn0&\sx0&\sn0&\sn0&\sn0&\tc{$i$}&\sn0&\tc{$i$}&\sn0&\sn0&\sn0&\tc{$-i$}&\sn0&\sn0&\sn0&\sx0&\sn0&\sx0&\sn0&\sn0&\sn0&\sx0&\sn0&\sx0&\sn0&\sn0&\sn0&\sx0&\sn0&\sx0&\sn0&\sx0&\sn0&\sn0&\sn0&\sx0&\sn0& \wire{$z\xor{s_5}$} \\
\end{tabular}

\end{center}
\caption{5-bit ripple carry adder written in terms of controlled
rotations.  The depth is 28.  Here a circled $i$ denotes a ``square
root of {\NOT}''; i.e., a rotation by $\pi/2$.  A circled $-i$ denotes
the inverse operation.}
\label{ripple-cr-fig}
\end{sidewaysfigure}

It is well-known that a Toffoli gate can be built from five
controlled rotations.  One might thus expect the controlled-unary
depth of our ripple-carry adder to be $10n + O(1)$.  In fact, the
Toffolis can be overlapped; the depth is only $6n-2$.  An example with
$n = 5$ is depicted in Figure~\ref{ripple-cr-fig}.

We can also consider the cost of adding a classical quantity
to a quantum quantity.  We have some $n$-bit number in our quantum memory,
and we wish to add a fixed $n$-bit number (known at compile time).  Our
ripple-carry adder does not become any simpler in this setting; we still
need to use $n$ quantum bits to store the classical addend.  On the other
hand, the transform adder benefits greatly:\ the classical information
need not be stored in quantum memory, and the controlled rotations are
replaced with fixed and known rotations.  In this setting, the transform
adder seems superior.

\bibliography{adder}
\bibliographystyle{amsplain}

\end{document}